\documentclass[%
 reprint,
 amsmath,amssymb,
 aps,
 superscriptaddress,
]{revtex4-1}

\usepackage{graphicx}
\usepackage{dcolumn}
\usepackage{bm}
\usepackage{color}

\usepackage{multirow}

\begin{document}

\preprint{APS/123-QED}

\title{Frustrated Spin One On A Diamond Lattice}

\author{J. R. Chamorro}
\affiliation{Department of Chemistry, The Johns Hopkins University, Baltimore MD 21218 USA}
\affiliation{Institute for Quantum Matter, Department of Physics and Astronomy, The Johns Hopkins University, Baltimore MD 21218 USA}

\author{L. Ge}
\affiliation{School of Physics, Georgia Institute of Technology, Atlanta, GA 30332, USA}

\author{J. Flynn}
\affiliation{Department of Chemistry, Oregon State University, Corvallis, OR, 97331, USA}

\author{M. A. Subramanian}
\affiliation{Department of Chemistry, Oregon State University, Corvallis, OR, 97331, USA}

\author{M. Mourigal}
\affiliation{School of Physics, Georgia Institute of Technology, Atlanta, GA 30332, USA}
 
\author{T. M. McQueen}
\email{mcqueen@jhu.edu}
\affiliation{Department of Chemistry, The Johns Hopkins University, Baltimore MD 21218 USA}
\affiliation{Institute for Quantum Matter, Department of Physics and Astronomy, The Johns Hopkins University, Baltimore MD 21218 USA}
\affiliation{Department of Materials Science and Engineering, The Johns Hopkins University, Baltimore MD 21218 USA}

\begin{abstract}

We report the discovery of a spin one diamond lattice in NiRh$_2$O$_4$. This spinel undergoes a cubic to tetragonal phase transition at $\it{T}$ = 440 K that leaves all nearest neighbor interactions equivalent. In the tetragonal phase, magnetization measurements show a Ni$^{2+}$ effective moment of $p_{\textrm{eff}}$ = 3.3(1) and dominant antiferromagnetic interactions with $\Theta_{\mathrm{CW}}$ = -11.3(7) K. No phase transition to a long-range magnetically ordered state is observed by specific heat measurements down to $\it{T}$ = 0.1 K. Inelastic neutron scattering measurements on sub-stoichiometric NiRh$_{2}$O$_{4}$ reveal possible valence-bond behavior and show no visible signs of magnetic ordering. NiRh$_2$O$_4$ provides a platform on which to explore the previously unknown and potentially rich physics of spin one interacting on the diamond lattice, including the realization of theoretically predicted quantum spin liquid and topological paramagnet states.

\end{abstract}

\maketitle

The recognition that there exist multiple classes of insulators not adiabatically connected to each other has resulted in numerous discoveries. These include 2D and 3D topological insulators \cite{Brune2011, Fu2007, Hsieh2008, Zhang2009}, Dirac and Weyl semimetals \cite{Liu2014,Neupane2014,Yang2015,borisenko2016}, candidate hosts for Majorana fermions \cite{Nadj-Perge2014}, and candidate topological superconductors \cite{Sasaki2011, Polley2016, Arpino2014}. These experimental discoveries have, in turn, spurred significant theoretical efforts to apply the tools of topological classification to other areas, such as in systems where electron correlations are strong, including topological magnons \cite{Zhang2013,Chisnell2015} and topological paramagnets \cite{Wang2015}. 

In correlated magnetic systems, competing interactions between magnetic moments can lead to geometric magnetic frustration, due to the inability of the system to satisfy all pairwise interactions due to the geometry of the lattice. Since frustration prevents the emergence of a single low energy ground state, it enables a variety of exotic states of matter, such as valence bond solids, spin liquids, and chiral-spin and spin-ice materials \cite{Affleck1987, Matan2010, Pratt2011, Anderson1987, Bode2007, Grohol2005, Taguchi2001, Fennell2009}. Quantum magnets host one of the earliest realizations of topological matter: the Haldane chain comprising of antiferromagnetically interacting spin one ions in one dimension \cite{Haldane1983}, so it is natural to ask how the physics of Haldane chains evolves in the presence of competing interactions between magnetic moments.

Recent work has suggested that a frustrated diamond lattice of S = 1 ions may result in a structure containing fluctuating and interconnected Haldane-like chains, whose excitation spectrum is gapped but possesses topologically non-trivial edge states \cite{Affleck1987, Wang2015}. Such an arrangement may give rise to a topological state not electronic in nature, but rather magnetic: a topological paramagnet. Other work suggests that such a quantum magnet might instead host a quantum paramagnetic state \cite{Chen2017, Buessen2017}.

\begin{figure}[h]
\centering 
\includegraphics[width=9cm]{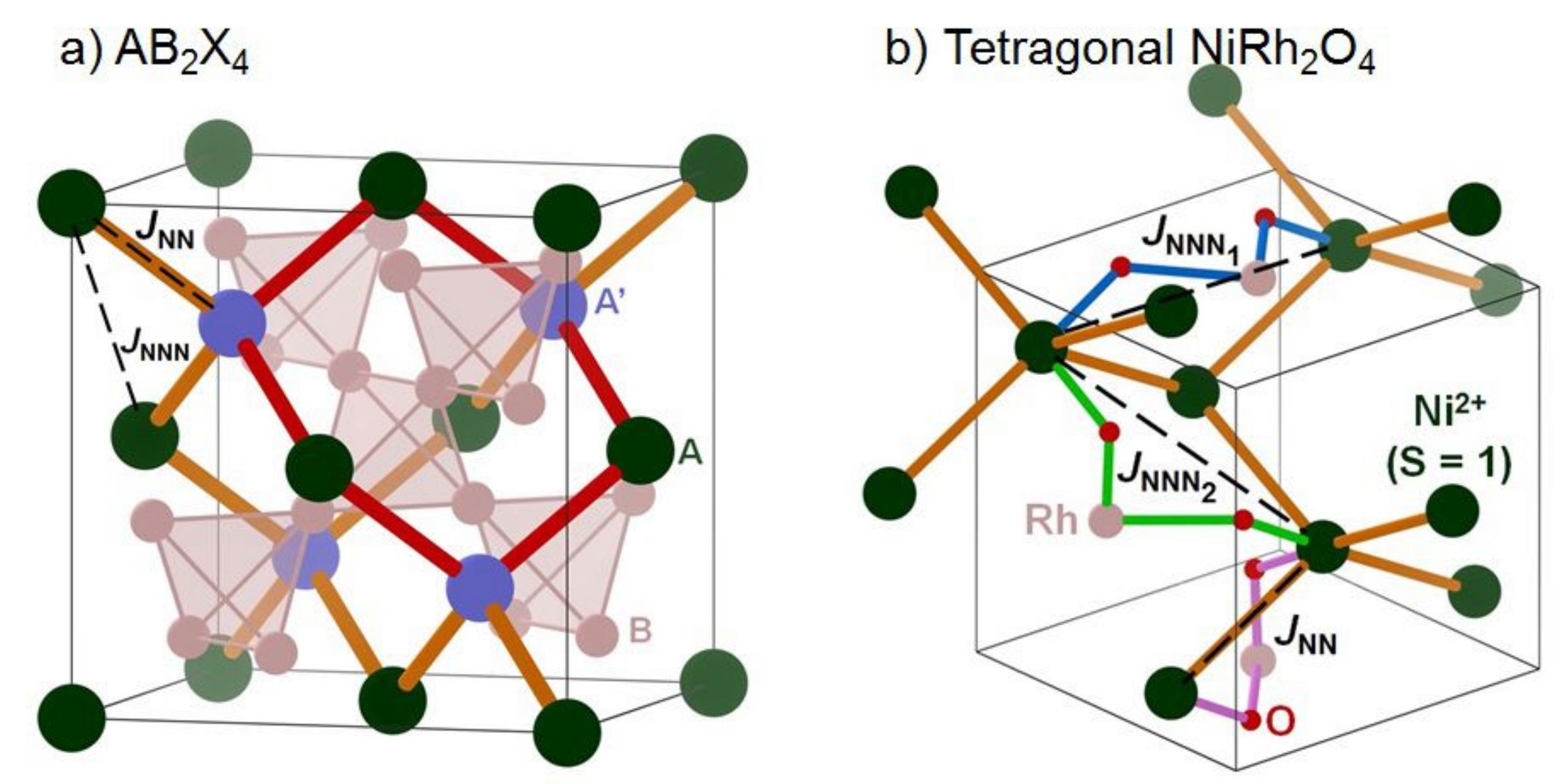}
{\caption{\textbf{(a)} The structure of a cubic AB$_2$X$_4$ spinel, consisting of a corner-sharing tetrahedral network of B-ions and a bipartite diamond lattice of A-ions. The diamond lattice is a 3D version of the honeycomb network (one hexagon highlighted). \textbf{(b)} NiRh$_2$O$_4$ is a realization of S = 1 on the diamond lattice, with non-magnetic B-ions (Rh$^{3+}$, low-spin d$^6$). Below \textit{T} = 440 K, NiRh$_2$O$_4$ is tetragonal, preserving equivalent NN interactions, but with two distinct NNN interactions. Possible superexchange pathways are shown.}}
\end{figure}

The diamond lattice can be found in the AB$_2$X$_4$ spinel structure type, best known for its frustrated pyrochlore lattice of B-site ions, as shown in \textbf{Fig. 1a}. The A-site diamond lattice is bipartite, composed of two interlacing face-centered-cubic (\textit{fcc}) sublattices, and can be viewed as the three-dimensional analogue of a honeycomb lattice. Within the diamond lattice, there are four nearest neighbor NN interactions between adjacent magnetic ions on separate fcc sublattices, and twelve next nearest neighbor NNN interactions between adjacent magnetic ions within each \textit{fcc} sublattice. As with the honeycomb lattice, a N\'eel ground state is expected in the presence of solely NN Heisenberg interactions \cite{Ge2017}. However, the N\'eel state is destabilized in the presence of NNN Heisenberg interactions that are at least $1/8$th as strong as the NN interactions, and a spiral spin liquid phase emerges \cite{Bergman2007}. No materials with S = 1 on the diamond lattice are known to date \cite{Wheeler2010,Krizan2015}.

Although there are many known chalcogenide spinel structures with magnetic A-site ions, none are \mbox{S = 1 \cite{Fritsch2004,Nilsen2015,Tristan2005,Shimizu2008,Zhang2006,Kojima2014}}. An ion with S = 1 on the A-site, such as Ni$^{2+}$ (d$^8$), would yield an S = 1 diamond lattice. However, the A-site in a spinel is tetrahedrally coordinated by chalcogen anions, and tetrahedrally coordinated Ni$^{2+}$ is exceptionally difficult to stabilize: crystal field stabilization energies strongly favor Ni$^{2+}$ on the B-site, which has octahedral coordination. Thus, virtually all nickel spinels are in fact inverse spinels, such as in M(Ni$_{0.5}$M$_{0.5}$)$_2$O$_4$ (NiM$_2$O$_4$, M= Ga, Al) \cite{Otero1985,Cooley1972}.

An exception to this pattern is NiRh$_2$O$_4$, which contains Ni$^{2+}$ on the A-site diamond lattice. Ni$^{2+}$ is forced onto the A-site by the very large octahedral crystal field stabilization energy gained by putting non-magnetic, low spin Rh$^{3+}$ (d$^6$) on the octahedrally-coordinated B-site. Previous reports on the synthesis and physical properties of NiRh$_2$O$_4$ are scant. They suggest that, at elevated temperatures, NiRh$_2$O$_4$ adopts the ideal cubic spinel structure, but that at $T \sim 380$ K it undergoes a cubic-to-tetragonal phase transition \cite{Blasse1963-1}, possibly associated with a Jahn-Teller (JT) distortion of the Ni$^{2+}$ ions \cite{Horiuti1964}. Previous studies have indicated the possibility of magnetic ordering at \textit{T}$_N$ = 18 K, however the rollover in the susceptibility is broad \cite{Blasse1963-2}. The local hyperfine field measured by $^{61}$Ni M\"ossbauer is very small, $\mu_0H$ = 2.5 T, indicative of a lack of magnetic ordering \cite{Gutlich1984}. These observations suggest that either only a small portion of the total moment is ordered, or that the entirety of the moment remains fluxional, similar to the behavior found for FeSc$_2$S$_4$ \cite{Fritsch2004,Plumb2016,Morey2016,Laurita2015}. Though the NN interactions ($J_{\mathrm{NN}}$) remain equivalent across the structural phase transition, the NNN interactions split into a set of four interactions within a plane ($J_{\mathrm{NNN_1}}$) and a set of eight interactions out of that plane ($J_{\mathrm{NNN_2}}$). The superexchange pathways for all three interactions are similar, based on -O-Rh-O- connectivity, shown in \textbf{Fig. 1b} implying that NN and NNN interactions can be of comparable magnitude.

Here we show that NiRh$_2$O$_4$ realizes S = 1 on a diamond lattice. We compare two samples of different stoichiometry, namely NiRh$_2$O$_4$ and Ni$_{0.96}$(Rh$_{1.90}$Ni$_{0.10}$)O$_4$. High resolution x-ray diffraction data confirm the complete occupancy of the A-site by Ni$^{2+}$ in the former, and neutron diffraction measurements confirm the stoichiometry in the latter. Specific heat measurements of NiRh$_2$O$_4$ show no signs of a phase transition to long range magnetic order down to \textit{T} = 0.1 K, and instead show a gradual loss of entropy over a wide temperature range. Magnetization data for NiRh$_2$O$_4$ indicate no magnetic ordering and a paramagnetic moment \mbox{$p_{\textrm{eff}}$ = 3.29(14)} with a Weiss constant $\Theta_\mathrm{{CW}}$ = -11.3(7) K, and the onset of spin glass behavior at $T \sim 6 $ K in Ni$_{0.96}$(Rh$_{1.90}$Ni$_{0.10}$)O$_4$. The frustration parameter, used as a qualitative measure of frustration in magnetic systems and defined as $f = |\Theta_\mathrm{{CW}}|/T_N$, is at least $f \geq 100$ in NiRh$_2$O$_4$, which is large in comparison to other known frustrated A-site spinel systems \cite{Kalvius2009,Roy2013}. Inelastic neutron scattering data set a stringent upper bound on any long-range order in Ni$_{0.96}$(Rh$_{1.90}$Ni$_{0.10}$)O$_4$ and provide evidence for dominant gapped excitations. Models fit to the data show potential valence bond behavior, as observed in other frustrated quantum magnets \cite{Sheckelton2012,Mourigal2014}. The neutron scattering data, in conjunction with the other measurements, indicate that NiRh$_2$O$_4$ is a candidate topological paramagnet.

\begin{figure} [h]
\includegraphics[width=7cm]{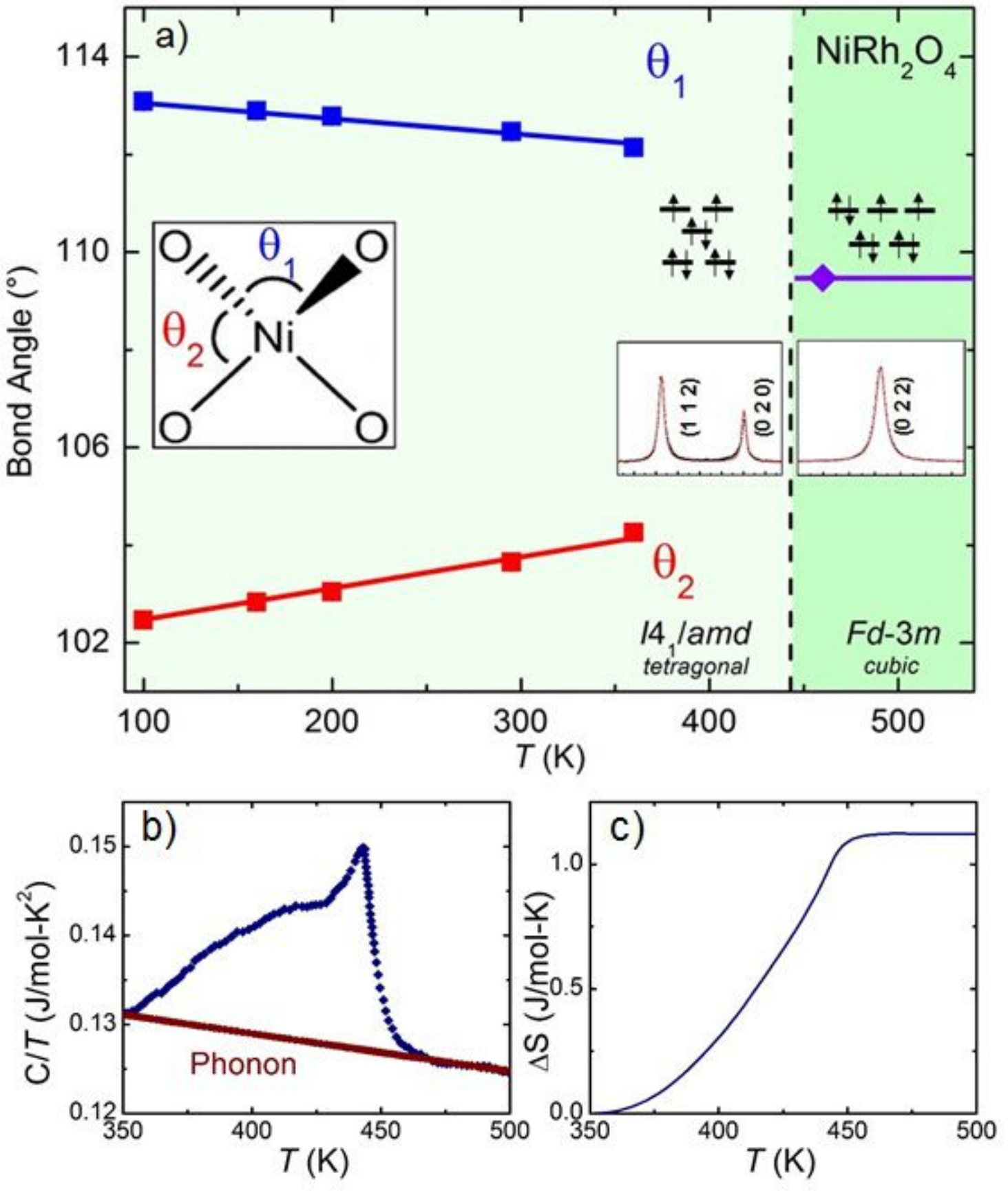}
\caption{Synchrotron XRD and DSC measurements show a cubic to tetragonal phase transition in NiRh$_2$O$_4$ at \textit{T} = 440 K. \textbf{(a)} Bond angles for each NiO$_4$ tetrahedron shows the phase transition corresponds to a JT distortion breaking some of the orbital degeneracy present in the cubic phase. \textbf{(b)} DSC measurements show a phase transition at $T = 440$ K. \textbf{(c)} The entropy loss is $\Delta$\textit{S}=0.13\textit{R}, significantly smaller than the 1.1\textit{R} required for a fully quenched orbital degree of freedom.}
\end{figure}

Polycrystalline NiRh$_2$O$_4$ was synthesized by sintering a stoichiometric mixture of green NiO (99.998\%) and hydrated Rh$_2$O$_3$ (99.99\%) under flowing oxygen at 1323 K for several days with intermediate regrindings. Ni$_{0.96}$(Rh$_{1.90}$Ni$_{0.10}$)O$_4$ was synthesized in a similar manner, without flowing oxygen. High resolution synchrotron x-ray diffraction data was collected on the former on 11-BM at the Advanced Photon Source at temperatures $T = 100-460$ K. Rietveld refinements reveal a cubic structure for the high temperature phase and a tetragonal structure for the low temperature phase; select results shown in \textbf{Fig. 2a} and \textbf{Table 1}. A small ($\sim0.2\%$) NiO impurity is included in all Rietveld refinements. The elevated R$_{wp}$ values in each fit can be ascribed to the peak shapes generated by the synchrotron data collection \cite{Cottingham2014}.

The high temperature phase has cubic symmetry with space group $Fd\bar{3}m$. This structure is in agreement with previous reports, with Ni$^{2+}$ in an ideal tetrahedral coordination by O$^{2-}$. Inclusion of anti-site Ni/Rh mixing and off-stoichiometry, such as excess Ni on Rh site, did not improve the quality of the refinement. The low temperature phase has tetragonal symmetry with space group $I{4_{1}}/amd$. The ${c_{cubic}}/{a_{cubic}} = {c_{tetragonal}}/{\sqrt{2}a_{tetragonal}}$ ratio, 1.05, indicates a large ($\sim 5\%$) departure from cubic symmetry. Initial attempts to fit the data to the previously reported tetragonal model were unsatisfactory and did not accurately capture the observed Bragg intensities. Further, the literature structures retain nearly perfect tetrahedral coordination of nickel by oxygen, showing only a large distortion around the Rh$^{3+}$ ions, unexpected on chemical grounds. Possible distortions of the unit cell for the tetragonal phase were explored using the online resource ISODISTORT \cite{Campbell2006}. An adequate fit to the data was obtained using a structure in which there is a buckling of the oxygen atoms in the (0 1 3) plane. This corresponds to a “pinching” of the NiO$_4$ tetrahedra along the \textit{c} axis, the bond angles of which are shown in \textbf{Fig. 2a}. The net result is that the NiO$_4$ tetrahedra become more linear with decreasing temperature. This naturally explains the observed change in \textit{c/a} ratio, in which the \textit{c} axis elongates and the \textit{a} and \textit{b} axes contract. It is also consistent with the previously proposed JT distortion, as the site symmetry of the Ni$^{2+}$ tetrahedra changes from a point group of $T_d$ to $D_{2d}$. 

To estimate the thermodynamic parameters of this phase transition, differential scanning calorimetry (DSC) measurements were carried out in heating mode using a TA instruments Q2000 DSC, as shown in \textbf{Fig. 2b}. The transition from cubic to tetragonal occurs at \textit{T} = 440 K, with an entropy loss of $\Delta$S = 0.13\textit{R}, as shown in \textbf{Fig. 2c}. This loss is significantly less than the $\Delta$S = \textit{R}ln(3) that would be expected if the JT distortion completely lifts the degeneracy of the orbital degrees of freedom. The transition temperature is also higher than the \mbox{\textit{T} = 380 K} previously reported, possibly due to an improvement in sample stoichiometry. Rietveld refinements of neutron diffraction data for Ni$_{0.96}$(Rh$_{1.90}$Ni$_{0.10}$)O$_4$ show the same structure as NiRh$_2$O$_4$, but with a slight degree of site mixing, as implied by the formula. Refinements for the synchrotron X-ray diffraction data at several temperatures on NiRh$_{2}$O$_{4}$ and the neutron powder diffraction data on Ni$_{0.96}$(Rh$_{1.90}$Ni$_{0.10}$)O$_4$ can be found in the S.I.

\begin{table}[h]
\begin{center}
\begin{tabular}{l*{3}{c}r}
 & \textit{T} = 100 K & \textit{T} = 460 K \\
\hline
Ni $B_{iso}$ & 0.0629 (1) & 0.6601 (1) \\
Rh $B_{iso}$ & 0.1595 (1)  & 0.6143 (1) \\
O $x$ & 0.7614 & \multirow{3}{*}{\Huge$\rbrace$ \hspace{2cm}}\\
O $y$ & 0.7500 (17) & 0.38255 (11)\\
O $z$ & 0.51682 (12) &  \\
O $B_{iso}$ & 0.115 (1)  &  0.925 (1)\\
\end{tabular}
\caption{ Structural analysis of cubic and tetragonal NiRh$_2$O$_4$. Tetrahedral $(T = 100$ K): $I{4_{1}}/amd$, $a = b = 5.88680 (1)$ \AA, $c = 8.71541 (2)$ \AA. Ni: \textit{4a} (1/2, 3/4, 3/8); Rh: \textit{8d} (1/2, 3/4, 3/4); O: \textit{16h} (0, \textit{y}, \textit{z}). $R_p = 12.9$; $R_{wp} = 15.9$; $\chi^2 = 1.3$. Cubic $(T = 460$ K): $Fd\bar{3}m$, $a = 8.47674 (1)$ \AA. Ni: \textit{8a} (0, 0, 0); Rh: \textit{16d} (5/8, 5/8, 5/8); O: \textit{32e} (\textit{x}, \textit{x}, \textit{x}). $R_p = 14.6$; $R_{wp} = 16.9$; $\chi^2 = 1.4$. All occupancies were held at unity in final refinements.}
\end{center}
\end{table}

\textbf{Fig. 3} shows the magnetic susceptibility of NiRh$_2$O$_4$ in the tetragonal phase, estimated as $\chi\sim$  M/H with magnetization measured using the ACMS option of a Quantum Design physical property measurement system (PPMS) with an applied field of $\mu_{0}H$ = 1.0T. The magnetic susceptibility of Ni$_{0.96}$(Rh$_{1.90}$Ni$_{0.10}$)O$_4$ is also shown, measured using the vibrating sample magnetometer (VSM) option. For NiRh$_2$O$_4$, nearly perfect Curie-Weiss behavior is observed, and these values agree with the slopes of M(H) curves measured at \textit{T} = 2 K and \textit{T} = 300 K, once the 0.2\% NiO contribution is taken into account. At low temperature, a slight antiferromagnetic deviation from Curie Weiss behavior is observed. No difference is observed in the zero-field cooled (ZFC) vs field cooled (FC) measurements. A Curie-Weiss analysis $\dfrac{1}{\chi-\chi_{0}}= \dfrac{1}{C}T-\dfrac{\Theta_\mathrm{{CW}}}{C}$ from 1.8 $\leq$ \textit{T} $\leq$ 300 K yields a Weiss temperature of $\Theta_\mathrm{{CW}}$ = $-11.3(7)$ K, indicative of net antiferromagnetic interactions. A paramagnetic moment of $p_{\textrm{eff}}$ = $\sqrt{8C}$  = 3.29(14) for Ni$^{2+}$ is observed. This value agrees with previous reports \cite{Gutlich1984}, and is intermediate between the spin-only value of 2.83$\mu_{B}$ and the spin-orbital value of 3.67$\mu_{B}$ based on L$_{\textrm{eff}}$ = 1 and \mbox{S = 1}. This is also in agreement with the DSC result, suggestive of a residual orbital contribution. A $\chi_{0}$ value of 0.00098 emu mol$^{-1}$ K$^{-1}$ Oe$^{-1}$ was used to account for temperature-independent contributions. Using the same $\chi_{0}$ value for the magnetization of Ni$_{0.96}$(Rh$_{1.90}$Ni$_{0.10}$)O$_4$, an effective moment of $p_{\textrm{eff}} = 2.81$ is obtained. This value is that of a pure spin S = 1 magnet, indicative of the absence of an orbital contribution to the moment. A positive \mbox{$\Theta_{CW}$ = $24.3$} K value indicates net ferromagnetic interactions.

\begin{figure}
\includegraphics[width=9.2cm]{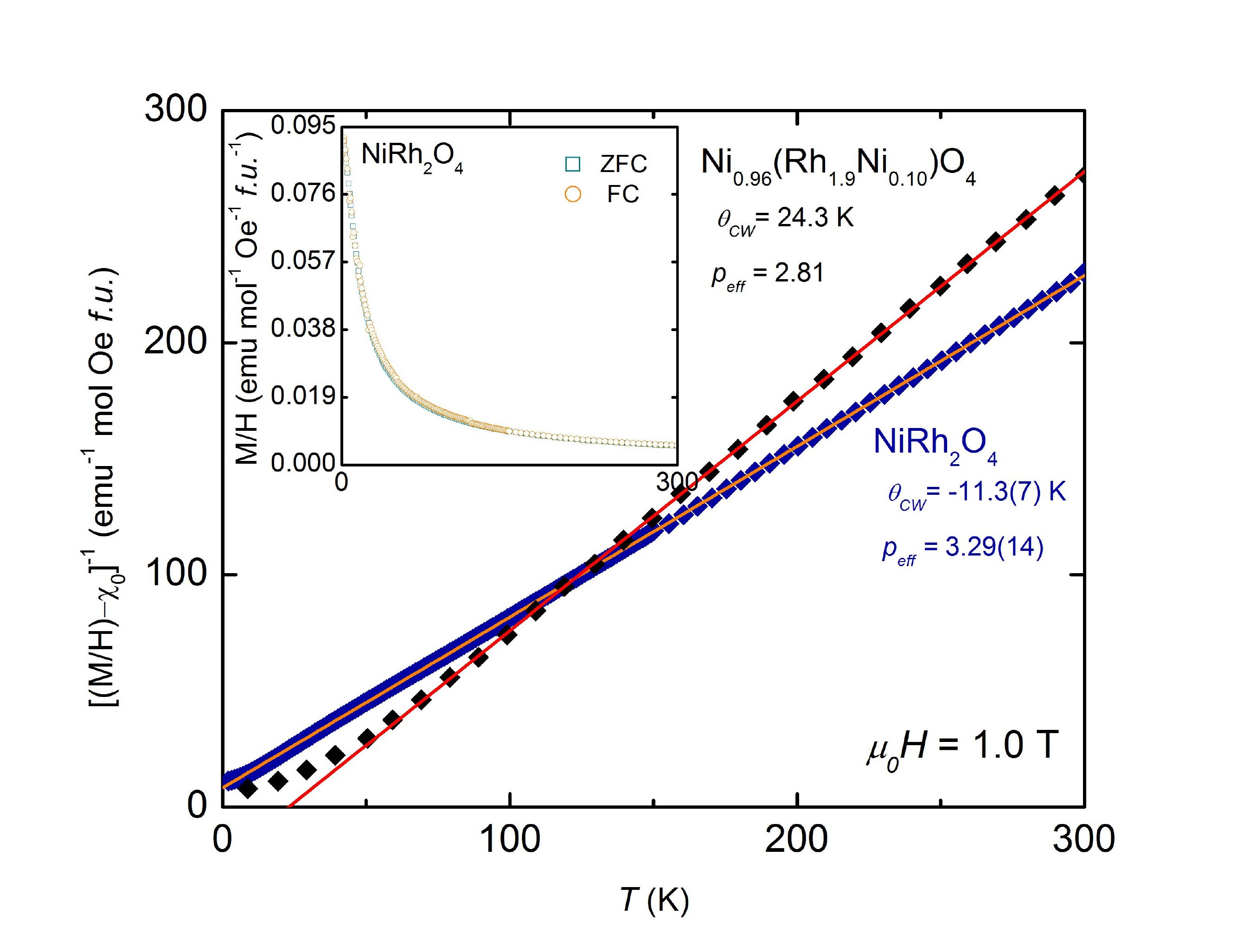}
\caption{Curie-Weiss analysis of magnetization measurements done on tetragonal NiRh$_2$O$_4$ and Ni$_{0.96}$(Rh$_{1.90}$Ni$_{0.10}$)O$_4$. The extracted effective moment for NiRh$_2$O$_4$ per Ni$^{2+}$ ion is greater than the spin only value of 2.83, unlike that for Ni$_{0.96}$(Rh$_{1.90}$Ni$_{0.10}$)O$_4$ which is effectively spin only. A Weiss temperature of $-11.3(7)$ K implies net antiferromagnetic interactions in NiRh$_2$O$_4$, and $24.3$ K for Ni$_{0.96}$(Rh$_{1.90}$Ni$_{0.10}$)O$_4$ implies ferromagnetic interactions. The inset shows the magnetization of NiRh$_{2}$O$_{4}$, which shows no difference in the zero field cooled (ZFC) and field cooled (FC) measurements.}
\end{figure}

Heat capacity data was collected at constant pressure using the PPMS heat capacity option, equipped with a dilution refrigerator. For both compounds, the resulting specific heat, \textbf{Fig. 4a}, shows no sharp anomalies that would be indicative of long range magnetic ordering or other phase transitions over a temperature from 0.1 $\leq$ \textit{T} $\leq$ 300 K. We estimated the phonon contribution to the specific heat through measurements of the non-magnetic analogue ZnRh$_2$O$_4$, synthesized by sintering stoichiometric amounts of ZnO and Rh$_2$O$_3$ in air at 1273 K for two days, and scaled by 0.97 to account for the difference in atomic mass between nickel and zinc. Though the structures for ZnRh$_2$O$_4$ and NiRh$_2$O$_4$ differ due to the JT distortion of the latter, the connectivity between corresponding atoms is the same, so the resulting phonon dispersions should be very similar. When this phonon contribution is subtracted, the excess specific heat of NiRh$_2$O$_4$ is attained. This shows two broad peaks, one with a maximum at \textit{T} = 1.77 K and the other with a maximum at \textit{T} = 33.7 K. The total excess, presumably magnetic, entropy is obtained by calculating the integral of the phonon-subtracted C/\textit{T}, as shown in \textbf{Fig. 4b}.  The entropy crosses $\Delta S$ = Rln(3) at \mbox{\textit{T} = 90 K}, and plateaus near $\Delta S$ = Rln(6) at \textit{T} = 250 K. This analysis is robust, with the total integrated entropy always remaining between Rln(5) and Rln(6) independent of the scaling method for the diamagnetic ZnRh$_2$O$_4$ analog. This exceeds the expected spin only entropy of Rln(3). On the other hand, the magnetic entropy of Ni$_{0.96}$(Rh$_{1.90}$Ni$_{0.10}$)O$_4$ does not reach Rln(6) and instead plateaus near Rln(3).

\begin{figure}[h]
\includegraphics[width=9cm]{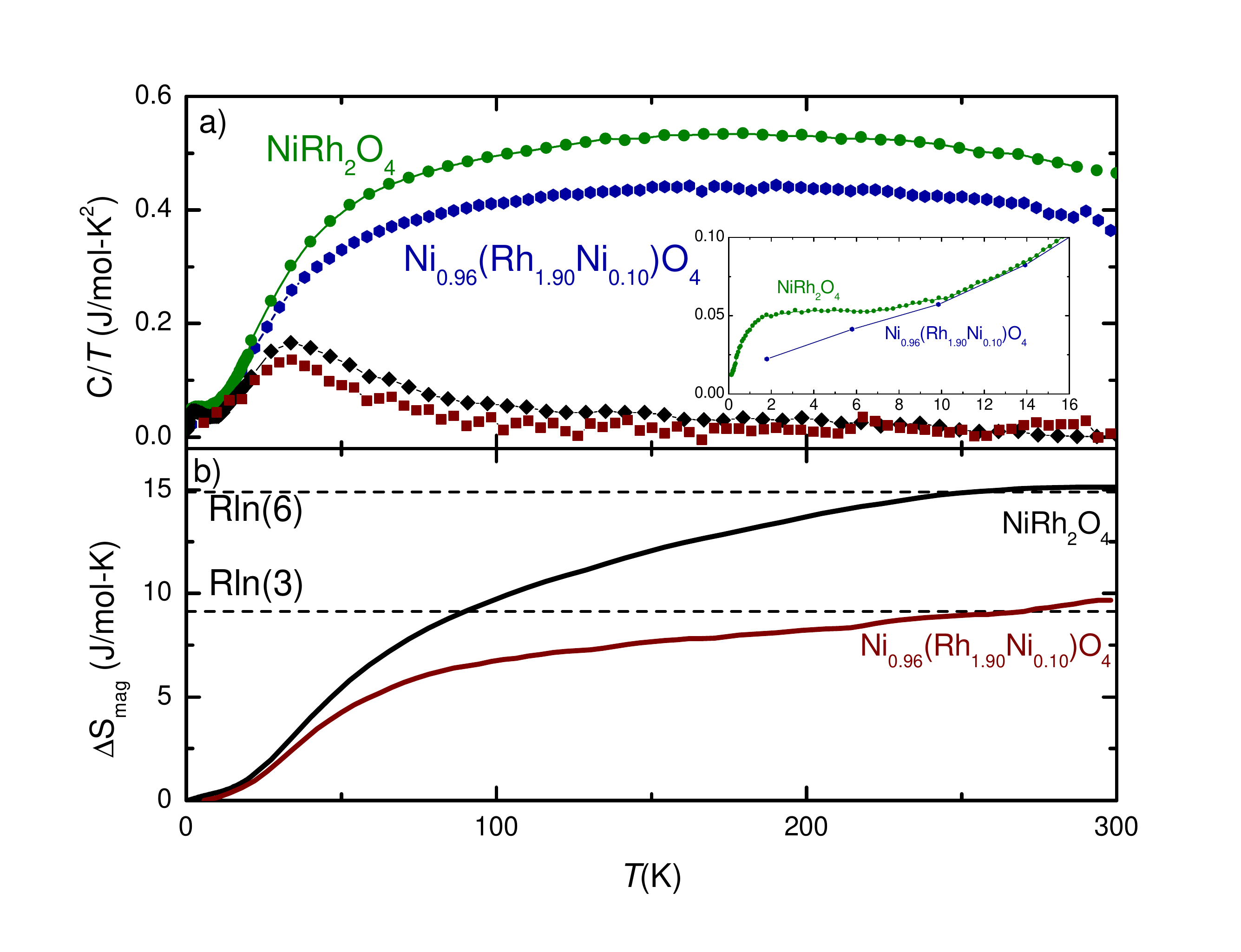}
\caption{\textbf{(a)} Specific heat over temperature for NiRh$_2$O$_4$ (circles) and  Ni$_{0.96}$(Rh$_{1.90}$Ni$_{0.10}$)O$_4$ ( from \textit{T} = 0.1 to 300 K. The phonon contribution (hexagons) is estimated from ZnRh$_2$O$_4$ and removed to leave the magnetic contribution for both (diamonds and squares, respectively). No sharp anomalies indicative of a phase transition are observed down to \textit{T} = 0.1 K. Inset shows the absence of the low temperature hump at \textit{T} = 1.77 K in Ni$_{0.96}$(Rh$_{1.90}$Ni$_{0.10}$)O$_4$. \textbf{(b)} Integration of C/\textit{T} yields the entropy of the magnetic component, which far exceeds the expected \textit{R}ln(3) spin only value for NiRh$_2$O$_4$, but not for Ni$_{0.96}$(Rh$_{1.90}$Ni$_{0.10}$)O$_4$.}
\end{figure}

The entropy is not recovered uniformly, however, for either compound. For stoichiometric NiRh$_{2}$O$_{4}$, there are three ranges: a low temperature hump that has a maxima at \textit{T} = 1.77 K, a higher temperature hump at \mbox{\textit{T} = 33.7 K}, and a continuum from 120 $\leq$ \textit{T} $\leq$ 300 K. The hump at \textit{T} = 1.77 K recovers an entropy of $\Delta$\textit{S} = 0.051\textit{R}, equivalent to $\sim 1\%$ of free S = 1/2 spins. This is too small to arise from a bulk phase transition and instead likely originates from vacancies, defect sites, or surface states. This could also originate from a nuclear contribution, such as that seen in other Ni compounds at similar temperature ranges \cite{Neilson2012}. This is further supported by a field dependence at a level commensurate with isolated spins (not shown).  The hump at \textit{T} = 33.7 K, present in both NiRh$_{2}$O$_{4}$ and Ni$_{0.96}$(Rh$_{1.90}$Ni$_{0.10}$)O$_4$, can be semi-quantitatively modeled as a two-level Schottky anomaly, given by: 

\begin{equation}
\footnotesize{\dfrac{C_\mathrm{{Schottky}}}{T} = OSF \cdot \Big(\dfrac{\Delta}{T}\Big)^2 \dfrac{e^{\small\dfrac{\Delta}{T}}}{\Bigg[1+e^{\small\dfrac{\Delta}{T}}\Bigg]^2} \cdot T^{-1}}
\end{equation}

Which, for NiRh$_{2}$O$_{4}$, corresponds to a difference between two energy levels of $\Delta$ = 116(3) K and an overall scale factor OSF = 1.87(7), corresponding to an entropy of $\Delta$\textit{S}=1.87(7)·Rln(2) = 1.3R. This is $\sim 75\%$ of the total excess entropy, leaving $\sim 22\%$ (0.4R) in the broad continuum from \textit{T} = 120 to 300 K. The $\Delta$\textit{S} = 1.3R entropy contained in the \textit{T} = 33.7 K hump is somewhat larger than the $\Delta$\textit{S} = \textit{R}ln(2S+1) = \textit{R}ln(3) = 1.1R spin entropy expected for spin one in the absence of orbital degrees of freedom. This is even excluding the excess entropy in the $T = 120-300 K$ range, and indicates a contribution from either orbital or phonon degrees of freedom (or both).

The inelastic neutron scattering (INS) experiment on Ni$_{0.96}$(Rh$_{1.90}$Ni$_{0.10}$)O$_4$ was carried out using the Fine-Resolution Fermi Chopper Spectrometer (SEQUOIA)
at Oak Ridge National Laboratory's Spallation Neutron Source. Four grams of powder sample were held in an aluminum can and measured at various temperatures from $4$ K to $300$ K with incident neutron energy $E_i = 80$ meV. The contribution from the empty can was removed during data reduction.

The INS intensity $I (Q, E) $ is presented in \textbf{Fig. 5} as a function of powder-averaged momentum-transfer {\it Q} and energy transfer {\it E}, where a strong dispersive mode centered around $E = 11$ meV [Fig. 5(a)-(b)] emerges at low temperature and persists at $T = 40$ K, above the spin glass point ($\sim 6$ K). The data clearly demonstrates that the dominant magnetic response is gapped. Cuts through the $T = 4$ K magnetic response [Fig. 5 (c)-(d)] show more detailed momentum and energy dependence of the excitations. Particularly, the constant-$Q$ cut over \mbox{$3$ meV $\le E \le 7$ meV} indicates very weak intensity peaked around $\{h,k,l\} = \{1,0,1\}$ and $\{0,1,1\}$, indicating any incipient magnetic order to be indexed with a propagation vector {\bf k}$_m$ = 0. No magnetic Bragg peaks can be observed in cuts at the elastic-line, putting a higher limit on any ordered moment of $\approx 0.1 - 0.2 \mu_B$.

Despite the lack of long range order, we attempted at modeling the excitations with a numerical implementation of linear spin-wave theory with a N\'eel groud state (corresponding to {\bf k}$_m$ = 0 as aforementioned) \cite{Petit2011, Toth2015}. The closest simulation, with $J_{\mathrm{NN}} = 2.6$~meV, \mbox{$J_{\mathrm{NNN}_1} = -0.3J_{\mathrm{NN}}$}, $J_{\mathrm{NNN}_2} = 0.12J_{\mathrm{NN}}$ and $\Delta = 1.1$, is displayed in Fig. 5 (c)-(d) (See supplementary information for details). While roughly matching the bandwidth, spin-wave theory fails to accurately capture the broadness in both momentum and energy although disorder effects may provide a possible explanation. More importantly, it produces multiple bands of spin-wave excitations, which contrasts with the unique dominant branch found in the measurements.

To better model the short range correlation, we apply the powder-average equal-time structure factor of valence bonds $\tilde{I}_{\mathrm{vb}}(Q) \propto r_0^2/6\left|F(Q)\right|^2 S_{\mathrm{vb}}(Q)$, where \mbox{$S_{\mathrm{vb}}(Q) = \sum\limits_i m_i^2 \left[ 1 - \sin(Qd_i) / (Qd_i) \right]/\mu_\mathrm{B}^2$} \cite{Mourigal2014} with the sum up to NNN$_2$, $r_0^2 = 0.539 \times 10^{-12}$ cm, $F(Q)$ is the magnetic form factor of Ni$^{2+}$, $d_{i}$ are the distances between corresponding neighbors, and $m_{i}^2$ is the squared magnetic moment per formula unit. The fits, shown in \textbf{Fig. 5 (c)} (orange solid curve), produces a good match with the data with $|m_{2}/m_{1}| = 0.42$ and $|m_{3}/m_{1}| = 0.53$. Overall, the inelastic neutron scattering suggests the possible presence of quantum-effects through an excitation spectrum resembling that of gapped valence-bond systems.

\begin{figure}[h!]
\includegraphics[width=0.46\textwidth]{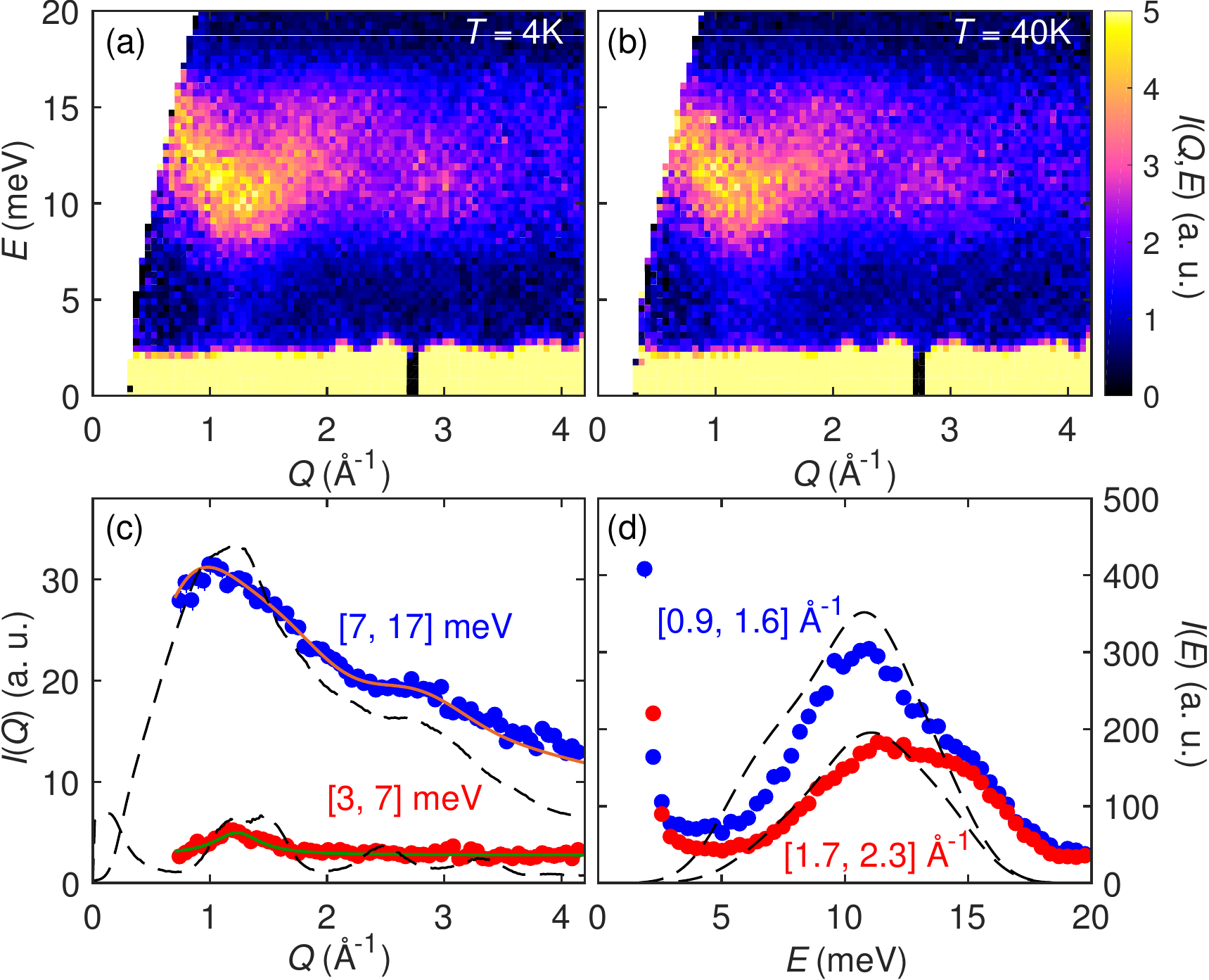}
\caption{Inelastic neutron scattering data on Ni$_{0.96}$(Rh$_{1.90}$Ni$_{0.10}$)O$_4$. \textbf{(a)-(b)} Scattering spectra $I(E,Q)$ at $T = 4$, $40$ K, respectively. \textbf{(c)-(d)} Constant-$Q$ (-$E$) cuts on various $E$ ($Q$) ranges of the $4$ K data. Dashed lines are linear spin-wave calculations. The solid, orange (color online) curve in \textbf{(c)} is the valence bond model and the solid, green one is a fit with the Lorentzian peak shape.}
\label{Fig 5}
\end{figure}

A plausible energy level analysis for Ni$^{2+}$ tetrahedra in NiRh$_2$O$_4$ that is consistent with our observations is shown in \textbf{Fig. 6}. The tetrahedral crystal field of Ni$^{2+}$ results in a splitting of the five orbitals into $e$ and $t_2$ sets. Assuming no $e - t_2$ electron excitations, placing eight electrons in the single particle levels gives rise to a series of multi-electron states, the lowest of which is a spin and orbital triplet $^{3}T_1$ \cite{Koester1975}. Spin-orbit coupling (SOC) and the JT distortion independently participate in splitting this $^{3}T_1$ state. SOC splits $^{3}T_1$ into four separate multi-electron states, based on their double group symmetries: lower energy $\Gamma_1$ and $\Gamma_4$, high energy $\Gamma_3$ and doubly degenerate $\Gamma_5$ \cite{Koster1963}. The JT distortion splits $^{3}T_1$ into two separate multi-electron states, $^3$E and $^{3}A_2$. $^3$E is a threefold degenerate manifold composed of $\Gamma_5$ and $\Gamma_1$. $^{3}A_2$ is six-fold degenerate and made up of $\Gamma_1$, $\Gamma_2$, $\Gamma_3$, $\Gamma_4$, and $\Gamma_5$. However, since the SOC and the JT distortion are on comparable energy scales, competition between these two interactions results in a mixing of multi-electron states. Using their double group symmetries, and given the differences in energy between the two sets of SOC energy levels, one arrives at a ground state manifold with a total of six states that account for the observed $\Delta$\textit{S} = Rln(6) in the magnetic specific heat in NiRh$_2$O$_4$, as shown highlighted in \textbf{Fig. 6b}.

\begin{figure*}[!ht]
\includegraphics[width=\textwidth]{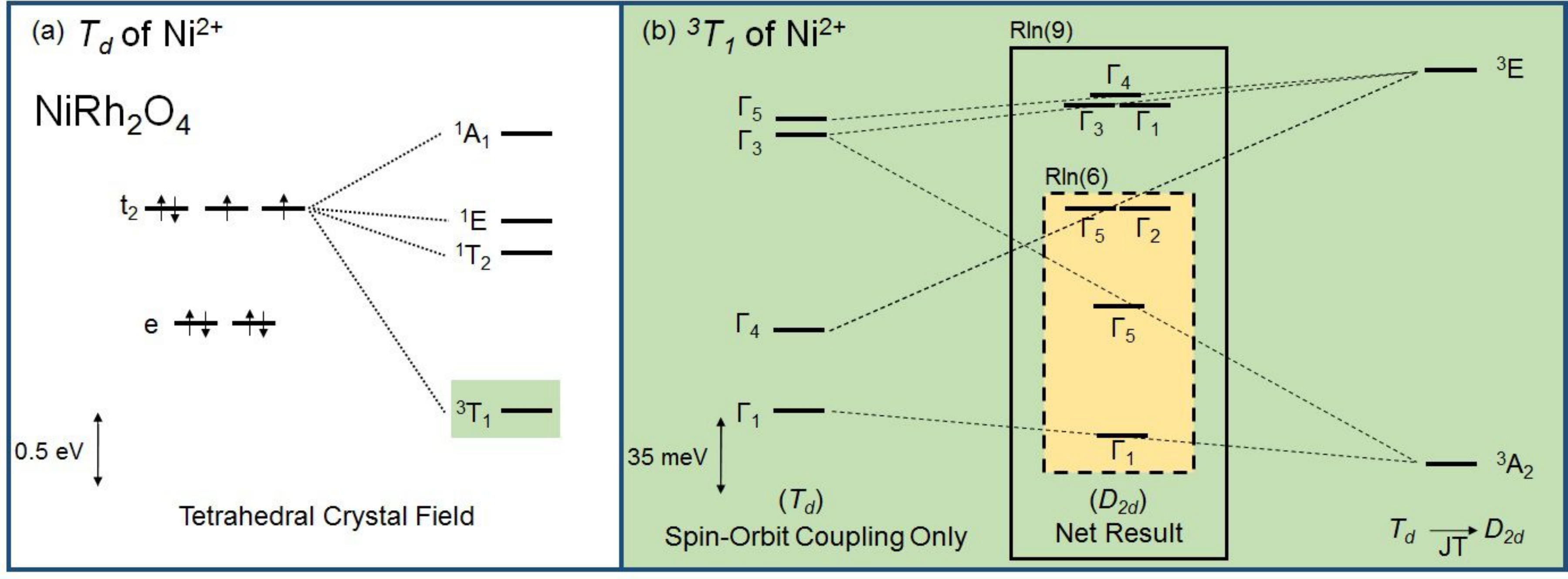}
\caption{Possible energy level analysis for Ni$^{2+}$ in NiRh$_2$O$_4$. \textbf{(a)} Tetrahedral crystal field results in a splitting of the five orbitals into $e$ and $t_2$ sets, with eight electrons that give rise to a multi-electron spin and orbital triplet $^3$\textit{T}$_1$ ground state. \textbf{(b)} SOC and JT interactions compete and further split the energy levels of the $T_d$ crystal field, consistent with the observed \textit{S} = \textit{R}ln(6) in the magnetic specific heat in the tetragonal phase.}
\end{figure*}

It is worth noting that the single-ion ground state predicted by a mix of spin orbit coupling and a Jahn-Teller distortion is a non-magnetic singlet ground state. The fairly small, antiferromagnetic $\Theta_\mathrm{{CW}}$ = $-11.3(7)$ K and the inelastic neutron scattering data point towards the possibility of such a ground state. The 11 meV bandwidth observed in our neutron scattering experiment seems to suggest that any valence bond behavior (such as any behavior tending towards spin liquid physics), or topologically non-trivial magnetic behavior (such as Haldane physics) is only apparent in the excitation spectrum. This is similar to $\alpha$-RuCl$_{3}$, a geometrically frustrated, honeycomb lattice material that shows spin liquid behavior in its high-energy excitations but not in the nature of its low temperature ground state.\cite{Banerjee2017,Wang2017,Baek2017}. Based on our putative analysis, the $\Gamma_{5}$ triplet single-ion excited state is a candidate for any exotic correlated behavior, as opposed to the single ion singlet $\Gamma_{1}$ ground state. 

The sensitivity of the physics of NiRh$_{2}$O$_{4}$ to the stoichiometry is also an indicator of the presence of unconventional strongly correlated behavior, as is seen in other systems with comparable, extreme sample sensitivity \cite{Fuhrman2017, Wen2017, Arpino2017}. While stoichiometric NiRh$_{2}$O$_{4}$ displays net mean field antiferromagnetic interactions, slightly off-stoichiometry Ni$_{0.96}$(Rh$_{1.90}$Ni$_{0.10}$)O$_4$ shows net ferromagnetic interactions, with the formation of a spin glass state at \textit{T} = 6 K. Though they both display a hump in the specific heat at \textit{T} = 33.7 K, their magnetic entropies plateau at different values. The difficulty in producing pure NiRh$_{2}$O$_{4}$ and the relative ease at which non-stoichiometric samples form likely explain the variation in physical properties in the literature. Furthermore, a comparison may be drawn to FeSc$_{2}$S$_{4}$, which hosts a fluxional, magnetically ordered ground state in proximity to a quantum critical point, as indicated by the suppression of antiferromagnetic ordering by the application of hydrostatic pressure \cite{Plumb2016}. The behavior of NiRh$_{2}$O$_{4}$ is not dissimilar and may therefore also be in proximity to a quantum critical point that may be accessed by some external parameter, such as pressure or magnetic field. 

In conclusion, we present magnetic susceptibility, specific heat, and synchrotron powder X-ray diffraction data on NiRh$_2$O$_4$ that confirm spin and orbital frustration despite undergoing a structural phase transition onset by a JT distortion. No phase transition to long range magnetic order is observed through specific heat and magnetization measurements down to \textit{T} = 0.1 K, in agreement with previous M\"ossbauer studies \cite{Gutlich1984}, indicating a significant degree of frustration. The entropy exceeds the $\Delta$\textit{S} = \textit{R}ln(3) value expected for an S = 1 system, indicating a remaining orbital or phonon contribution to the specific heat. Inelastic neutron scattering data demonstrate the presence of an excitation gap in sub-stoichiometric NiRh$_2$O$_4$ that likely persist in the fully stoichiometric counterpart. Whether this gap is due to single-ion anisotropy, valence-bond physics, or their interplay is a central question that warrants further investigation. Our work demonstrates that NiRh$_2$O$_4$ is a realization of S = 1 on the diamond lattice and is a platform for exploring the physics of such a frustrated, three-dimensional integer spin system. The extreme sample sensitivity warrants further study into the dynamic ground state and possible topological paramagnetism. 

The authors thank W. A. Phelan and S. Aubuchon (TA Instruments) for assistance with the differential scanning calorimetry measurements; J. A. M. Paddison and A. P. Ramirez for participation in early measurements and analysis; A. Huq and M. B. Stone for assistance in collecting data at ORNL;  and C. L. Broholm for useful discussions. The work at IQM was supported by the US Department of Energy, office of Basic Energy Sciences, Division of Materials Sciences and Engineering under grant DE-FG02-08ER46544. Use of the Advanced Photon Source at Argonne National Laboratory was supported by the U. S. Department of Energy, Office of Science, Office of Basic Energy Sciences, under Contract No. DE-AC02-06CH11357. The research at Oak Ridge National Laboratory's Spallation Neutron Source was sponsored by the U.S. Department of Energy, Office of Basic Energy Sciences, Scientific User Facilities Division. A portion of this research project was conducted using computational resources at the Maryland Advanced Research Computing Center (MARCC). The work at Oregon State University (M.A.S.) was supported by the National Science Foundation through Grant No. DMR- 1508527. JRC acknowledges support of the Greer Fellowship for Undergraduate Research. TMM acknowledges support of the David and Lucile Packard Science and Engineering Fellowship.

\bibliography{NiRh2O4.bib}

\end{document}